\documentclass{main}

\usepackage{graphicx}      
\usepackage{epstopdf}
\usepackage{subcaption}
\usepackage{algorithm}
\usepackage{algpseudocode}
\usepackage{optidef}
\usepackage{amsmath} 
\usepackage{amssymb}  
\usepackage{natbib}        
\usepackage{url}
\begin{document}
\begin{frontmatter}

\title{Data-Driven Control of Unknown Systems:\\A Linear Programming Approach\thanksref{footnoteinfo}} 

\thanks[footnoteinfo]{This research work was supported by the European Research Council (ERC) under the project OCAL, grant number 787845.\\$\copyright$ 2020 the authors. This work has been accepted to IFAC for publication under a Creative Commons Licence CC-BY-NC-ND.}

\author[First]{Alexandros Tanzanakis} 
\author[First]{John Lygeros} 

\address[First]{Department of Information Technology and Electrical Engineering, ETH Zurich, Switzerland, (e-mail: \{atanzana,jlygeros\}@ethz.ch).}

\begin{abstract}                
We consider the problem of discounted optimal state-feedback regulation for general unknown deterministic discrete-time systems. It is well known that open-loop instability of systems, non-quadratic cost functions and complex nonlinear dynamics, as well as the on-policy behavior of many reinforcement learning (RL) algorithms, make the design of model-free optimal adaptive controllers a challenging task. We depart from commonly used least-squares and neural network approximation methods in conventional model-free control theory, and propose a novel family of data-driven optimization algorithms based on linear programming, off-policy Q-learning and randomized experience replay. We develop both policy iteration (PI) and value iteration (VI) methods to compute an approximate optimal feedback controller with high precision and without the knowledge of a system model and stage cost function. Simulation studies confirm the effectiveness of the proposed methods.
\end{abstract}


\end{frontmatter}

\section{INTRODUCTION}
Reinforcement Learning (RL) bridges the gap between model-based and model-free control. This is accomplished by optimizing policies of (possibly) unknown dynamical systems with the goal of maximizing or minimizing a long-term reward function. The derivation of the Q-learning algorithm \citep{Watkins:24}, \citep{Bradtke:2}, along with Approximate Dynamic Programming (ADP) methods such as neurodynamic programming \citep{Bertsekas:32} were among the first approaches that dealt effectively with the problem of model-free optimal adaptive control.

To address the challenges associated with solving the Bellman and Hamilton-Jacobi-Bellman equations in model-based and model-free optimal control \citep{Lewis:1}, reliable ADP methods have been developed \citep{Powell:4}. These equations can be approximately solved by utilizing a family of iterative methods known as Policy Iteration (PI) and Value Iteration (VI) \citep{Bertsekas:7}, \citep{Lewis:8}.

In the Q-learning setting, an Actor-Critic framework \citep{Kiumarsi:5}, \citep{Konda:27} is commonly employed to approximate the Q-function and control policy with appropriate parametric function models (i.e. models with a priori fixed number of basis elements). For the computation of an approximate control policy, two fundamental learning schemes are used. In on-policy learning \citep{Wei:9}, \citep{Kiumarsi:10}, a single control policy is used for both generation of training data from the system and for online policy evaluation and improvement. In off-policy learning \citep{Li:14}, \citep{Li:11}, a behavior control policy is applied to the system and is responsible for the generation of training data samples, while a target control policy is iteratively evaluated and updated online. In the off-policy setting, a highly promising approach called Experience Replay \citep{Adam:15}, \citep{Liu:16}, \citep{Zha:17} is often employed to resolve the critical sample inefficiency issue of many learning algorithms, where real data samples are gathered and used only in a specific learning iteration, after the end of which they are discarded. PI and VI can be implemented in both schemes by applying either least-squares or neural network approximation methods. 

The Linear Programming (LP) approach to ADP is an alternative, model-based optimization paradigm to approximate the Value function \citep{Lerma:28}, \citep{Wang:21} or the Q-function \citep{Beuchat:20}, \citep{Cogill:22} and control policy, with rigorous theoretical guarantees on the approximation quality and online performance. For the model-free setting, \citep{Banjac:23} proposed an on-policy Q-learning-based PI LP algorithm for discounted state-feedback regulation of deterministic linear time-invariant (LTI) systems, utilizing a heuristic based on support constraints. 

Here, we propose a novel family of off-policy Q-learning-based LP algorithms for reliable discounted state-feedback regulation of general unknown deterministic discrete-time systems. The PI and VI methods are reformulated as data-driven, finite-dimensional linear programs, which proceed with the computation of an approximate optimal feedback controller without the knowledge of a system model and stage cost function. As a consequence, unlike the method proposed in \citep{Banjac:23}, our methods inherit the convergence guarantees of the standard PI and VI algorithms. To cope with the sample exploitation problem, we utilize a simple yet highly effective off-policy learning scheme called Randomized Experience Replay. To the best of our knowledge, this the first work in the related LP literature that provides a unified approach for effective model-free control of both deterministic discrete-time LTI and nonlinear systems.

The rest of this paper is organized as follows. The technical preliminaries are derived in Section $2$. The proposed family of data-driven optimization algorithms is presented and discussed in Section $3$. Simulation studies are carried on Section $4$ and conclusions are given in Section $5$.

\textbf{Notation.} $\mathbb{N}$ denotes the set of natural numbers excluding $0$. $\mathbb{R}$ defines the set of real numbers, while $\mathbb{R}_{+}$ the set of real numbers which are greater or equal than $0$. $\mathbb{S}^{n}$ denotes the set of $n\times n$ symmetric matrices, while $\mathbb{S}_{++}^{n}$ the set of $n \times n$ symmetric positive definite matrices. $I_{n\times n}$ denotes an $n\times n$ identity matrix. $tr(A)$ defines the trace of a matrix $A\in \mathbb{R}^{n\times n}$.
\section{TECHNICAL PRELIMINARIES}
\subsection{The problem of discounted state-feedback regulation}
Consider a deterministic discrete-time system of the form
\begin{equation}
x_{k+1} = f(x_k,u_k),
\end{equation}
where $x_k\in \mathcal{X}\subseteq \mathbb{R}^{n}$, $u_k\in \mathcal{U}\subseteq \mathbb{R}^{m}$ and $f:\mathcal{X}\times \mathcal{U}\rightarrow \mathcal{X}$ the model dynamics at time step $k$. The main objective of discounted state-feedback regulation is the computation of a control policy $\mu:\mathcal{X}\rightarrow \mathcal{U}$ that minimizes the cost
\begin{equation}
J^{\mu}(x_0)= \sum_{k=0}^{\infty}\gamma^{k}l\big(x_k,\mu(x_k)\big),
\end{equation}
where $\gamma \in (0,1)$ is the discount factor and $l:\mathcal{X}\times \mathcal{U}\rightarrow \mathbb{R}_{+}$ is the stage cost. To ensure the problem is well-posed, we assume that there exists a policy $\mu$ such that $J^{\mu}(x)<+\infty$ for all $x\in \mathcal{X}$. We are interested in the case where the functions $l$ and $f$ are not known, but their values can be observed for specific instances of $x$ and $u$, for example by sampling, simulating, or experimenting with the system. This is the typical setting in Q-learning.

\subsection{The optimal Q-function}
The optimal policy that minimizes $J^{\mu}$ can be computed if one has access to the optimal Q-function
\begin{equation}
Q^{\star}(x,u) = l(x,u) + \inf_{\mu} \sum_{k=1}^\infty \gamma^k l(x_k,\mu(x_k)).
\end{equation}
It can be shown that $Q^{\star}$ satisfies the Bellman equation
\begin{equation}
Q^{*}(x,u)=\underbrace{l(x,u)+\gamma \min_{v} Q^{*}\big(x',v\big)}_{FQ^{*}(x,u)},
\end{equation} 
where $x'=f(x,u)$ and $F$ is the Bellman operator for Q-functions. $F$ can be shown to be monotone and contractive \citep{Cogill:22}. Once $Q^{\star}$ is available, the optimal policy can then be computed by 
\begin{equation}
\mu^{*}(x) = \underset{v}{\mathrm{argmin }}Q^{*}(x,v).
\end{equation}
To obtain the linear programming formulation, one starts by relaxing the Bellman equation (4) to the Bellman inequality \citep{Beuchat:20}, \citep{Cogill:22}. An exact LP-based reformulation of (4) is then given by
\begin{equation}
\begin{aligned}
& \underset{Q\in \mathcal{F}(\mathcal{X},\mathcal{U})}{\text{max}} 
& & \int_{\mathcal{X}\times \mathcal{U}}Q(x,u)c\big(d(x,u)\big)\\
& \text{s.t.}
& &Q(x,u) \leq l(x,u)+ \gamma Q\big(x',v\big)\\
& & &\forall (x,u,v)\in \mathcal{X}\times \mathcal{U}^{2},\\
\end{aligned}
\end{equation} 
where $\mathcal{F}$ is the space of measurable functions bounded in an appropriate norm.
 
\textbf{Lemma 1 \citep{Beuchat:20}, \citep{Cogill:22}.} If $Q^{\star}\in \mathcal{F}(\mathcal{X},\mathcal{U})$, then the maximizer of (6) is identical to the solution of (4), for $c$ almost all $(x,u)\in \mathcal{X}\times \mathcal{U}$.

An important condition for the equivalence between (4) and (6) is that $\mathcal{F}(\mathcal{X},\mathcal{U})$ contains the optimal Q-function, i.e. the inequalities of problem (6) can be satisfied with equality. The quantity $c(\cdot,\cdot)$ is called state-action relevance weight. State-action relevance weights are finite measures and typically allocate a positive mass to all open subsets of $\mathcal{X}\times \mathcal{U}$ \citep{Beuchat:20}. 

The solution of (6) is intractable in general. The main difficulties are summarized as follows \citep{Beuchat:20}:
\begin{itemize}
\item[i)] The dimension of the space $\mathcal{F}(\mathcal{X},\mathcal{U})$ can be high (if $\mathcal{X}$ and $\mathcal{U}$ are finite) or even infinite (if they are not, the case of interest here).
\item[ii)] The number of inequality constraints of optimization problem (6) can be high or infinite.
\item[iii)] For an arbitrary $Q^{*}\in \mathcal{F}(\mathcal{X},\mathcal{U})$, the control policy calculation (5) may be intractable.
\end{itemize} 
These difficulties describe the curse of dimensionality issue in ADP for our problem setting. To tackle difficulty i), we construct a restricted function space $\hat{\mathcal{F}}(\mathcal{X}\times \mathcal{U})$ \citep{Beuchat:20}, \citep{Wang:21} of linear combinations of basis functions $\hat{Q}_{j}(\cdot,\cdot),j=1,\ldots,K$,
\begin{equation}
\hat{\mathcal{F}}(\mathcal{X}\times \mathcal{U})=\{Q(\cdot,\cdot)|Q(x,u) =\alpha^{T}\hat{Q}(x,u)\},
\end{equation}
where $\alpha \in \mathbb{R}^{K}$ and $\hat{Q}(x,u)=\begin{bmatrix} \hat{Q}_1(x,u),\ldots,\hat{Q}_K(x,u)\end{bmatrix}:\mathcal{X}\times \mathcal{U}\rightarrow \mathbb{R}^{K}$.
An approximate solution of (4) can then be computed by solving
\begin{equation}
\begin{aligned}
& \underset{Q\in \hat{\mathcal{F}}(\mathcal{X}\times \mathcal{U})}{\text{max}} 
& & \int_{\mathcal{X}\times \mathcal{U}}Q(x,u)c\big(d(x,u)\big)\\
& \text{s.t.}
& &Q(x,u) \leq l(x,u)+ \gamma Q\big(x',v\big)\\
& & &\forall (x,u,v)\in \mathcal{X}\times \mathcal{U}^{2}.\\
\end{aligned}
\end{equation} 
The optimizer $\hat{Q}^{\star}$ of (8) defines a policy
\begin{equation}
\hat{\mu}^{*}(x)= \underset{v}{\mathrm{argmin }}\hat{Q}^{*}(x,v).
\end{equation}
The hope is that, if we choose the space $\hat{\mathcal{F}}$ well enough, $\hat{Q}^{\star}$ will be a good approximation of $Q^{\star}$, so hopefully the performance of $\hat{\mu}^{\star}$ will be similar to that of $\mu^{\star}$.

To tackle difficulty (ii), $\mathcal{\hat{F}}(\mathcal{X},\mathcal{U})$ must be constructed based on suitable basis function elements, according to a particular application setting. For example, one can use quadratic functions and the S-procedure or polynomial optimization methods to derive a tight approximation of the infinite inequality constraints in (6). Likewise, one may want to restrict attention to basis functions that are convex in $u$ to deal with difficulty iii).

In general, the solution of (8) will also depend on the choice of the state-action relevance weight $c(\cdot,\cdot)$. The work of \citep{Beuchat:26} proposed solving the approximate LP problem (8) for multiple realizations of $c(\cdot,\cdot)$ and then computing a pointwise maximum over all derived solutions.  

We note that in the case that $Q^{\star} \in \mathcal{\hat{F}}(\mathcal{X}\times \mathcal{U})$, then the solution of the approximate LP problem (8) is $Q^{\star}$, as long as $c(\cdot,\cdot)$ allocates positive mass to all open subsets of $\mathcal{X}\times \mathcal{U}$. However, the standard LP approach to ADP requires both the system dynamics (1) and stage cost function to be known. In the following sections we will show how to bypass this requirement by deriving novel, data-driven LP variants of the well-known PI and VI methods.
\subsection{PI and VI in Q-learning}
In RL/ADP, the optimal Q-function and control policy are approximated online. In Q-learning, the Bellman equation (4) can be approximately solved using an iterative family of methods called PI and VI \citep{Bertsekas:7}, \citep{Lewis:8}. PI requires an initial control policy $\mu^{0}(x)$ such that $J^{\mu^0}(x) <\infty$ for all $x \in \mathcal{X}$. At iteration $i\geq 0$, it proceeds with the following two successive steps until convergence of the Q-function:
\begin{enumerate}
\item[i)] \textbf{PI - Policy Evaluation Step:} Solve for $Q^{i}$,
\begin{equation}
Q^{i}(x,u)=l(x,u)+\gamma Q^{i}\big(x',\mu^{i}(x')\big).
\end{equation}
\item[ii)] \textbf{PI - Policy Improvement/Update Step:}
\begin{equation}
\mu^{i+1}(x) = \underset{v}{\mathrm{argmin }}Q^{i}(x,v).
\end{equation} 
\end{enumerate}
It can be shown that PI provides non-increasing, monotone convergence to the optimal Q-function and control policy for both undiscounted and discounted optimal control problems. PI provides fast convergence to the optimal Q-function and control policy, although it requires an initial stabilizing policy, which is challenging in many cases to compute \citep[Chapter 4]{Wei:29}, \citep{Luo:30}, \citep{Heydari:6}.

VI can be initialized with an arbitrary $Q^{0}(x,u)\geq 0$ and control policy $\mu^{0}(x)$. At iteration $i\geq 0$, it proceeds with the following two steps until convergence of the Q-function:
\begin{enumerate}
\item[i)] \textbf{VI - Policy Evaluation Step:} Solve for $Q^{i+1}$,
\begin{equation}
Q^{i+1}(x,u)= l(x,u)+\gamma Q^{i}\big(x',\mu^{i}(x')\big).
\end{equation}
\item[ii)] \textbf{VI - Policy Improvement/Update Step:}
\begin{equation}
\mu^{i+1}(x) = \underset{v}{\mathrm{argmin }}Q^{i+1}(x,v).
\end{equation} 
\end{enumerate}
VI also provides theoretical guarantees related to monotonicity and convergence to the optimal Q-function and control policy. However, the convergence speed of VI is much slower than the one of PI \citep{Luo:31}, \citep{Heydari:6}. Finally, we note that the policy evaluation step for VI (12) is just a simple recursion and not an equation as in PI (10) .

\section{THE PROPOSED Q-PI-LP AND Q-VI-LP ALGORITHMS}
In this section, we present a novel family of data-driven, off-policy Q-learning based optimization algorithms, called Q-PI-LP and Q-VI-LP. Algorithms $1$ and $2$ show the proposed methods. 

In the off-policy setting, two types of control policies are defined: $i)$ a behavior policy, $a$, which is used for training data generation and $ii)$ a target policy, $\mu(x)$, which is successively evaluated and improved. To avoid confusion, we refer to the policy evaluation and improvement steps for PI and VI as target policy evaluation and improvement.
\begin{algorithm}
        \caption{Q-PI-LP algorithm.}\label{euclid}
        \begin{algorithmic}[1]
            \State Select threshold parameter $\epsilon >0$ and buffer size $N \in \mathbb{N}$.
            \State Experience Replay Buffer: Construct buffer $\mathcal{B}=\{(x_b,a_b,y_{b},l_b)\}_{b=1}^{N}$, where $(x_b,a_b)$ are random state-behavior policy sample pairs, $y_{b}$ is computed by applying $(x_b,a_b)$ to the unknown system (1) and $l_b$ is the measurement of the resulting stage cost.
            \State Pick $\hat{\mu}^{0}(x)$ such that $J^{\hat{\mu}^0}(x) <\infty$ for all $x \in \mathcal{X}$.
             \State Set $i=0$.
             \State Solve the LP problem,
\begin{align}
& \underset{Q^{i}\in \hat{\mathcal{F}}(\mathcal{X}\times \mathcal{U})}{\text{max}} 
& & \int_{\mathcal{X}\times \mathcal{U}}Q^{i}(x,a)c\big(d(x,a)\big) \nonumber\\
& \text{s.t.}
& &Q^{i}(x_b,a_b) \leq l_b+ \gamma Q^{i}\big(y_b,\hat{\mu}^{i}(y_b)\big) \nonumber\\
& & & \text{for }b=1,\ldots,N.
\end{align} 
\State Update $\hat{\mu}^{i+1}(x)=\underset{v}{\mathrm{argmin}}Q^{i}(x,v)$.
\State If $i\geq 1$ and $|Q^{i}(x_b,a_b)-Q^{i-1}(x_b,a_b)|\leq \epsilon$ for all $b$, then terminate; else set $i=i+1$, go to Step $5$ and continue.
        \end{algorithmic}
    \end{algorithm}

\begin{algorithm}
        \caption{Q-VI-LP algorithm.}\label{euclid}
        \begin{algorithmic}[1]
            \State Select threshold parameter $\epsilon >0$ and buffer size $N \in \mathbb{N}$.
            \State Experience Replay Buffer: Construct Experience Replay Buffer as in Algorithm 1.
            \State Choose $Q^0(x,a)\geq 0$ arbitrary.
             \State Compute $\hat{\mu}^{0}(x)=\underset{v}{\mathrm{argmin}}Q^{0}(x,v)$. 
             \State Set $i=0$.
             \State Solve the LP problem,
\begin{align}
& \underset{Q^{i+1}\in \hat{\mathcal{F}}(\mathcal{X}\times \mathcal{U})}{\text{max}} 
& & \int_{\mathcal{X}\times \mathcal{U}}Q^{i+1}(x,a)c\big(d(x,a)\big) \nonumber \\
& \text{s.t.}
& &Q^{i+1}(x_b,a_b) \leq l_b+ \gamma Q^{i}\big(y_b,\hat{\mu}^{i}(y_b)\big)\nonumber \\
& & & \text{for }b=1,\ldots,N.
\end{align}
\State Update $\hat{\mu}^{i+1}(x)=\underset{v}{\mathrm{argmin}}Q^{i+1}(x,v).$
\State If $|Q^{i+1}(x_b,a_b)-Q^{i}(x_b,a_b)|\leq \epsilon$ for all $b$, then terminate; else set $i=i+1$, go to Step $6$ and continue.
        \end{algorithmic}
    \end{algorithm}
\subsection{Randomized Experience Replay (RER)}
An important challenge in model-free optimal adaptive control is the satisfaction of the persistance of excitation (PoE) condition \citep{Tao:12}. PoE is required to ensure optimal parameter convergence and is generally guaranteed by designing appropriate probing noise, which is added to the control policy. However, in on-policy learning, the use of such probing noise can bias solutions \citep{Li:11}. 

Off-policy learning provides three convenient ways to reduce or even eliminate any bias of solutions under PoE \citep{Luo:19}:
\begin{enumerate}
\item[i)] Construction of an offline batch of training data of a sufficiently large size $N$. This batch of data will be repeatedly used in the learning phase of an underlying off-policy algorithm. This family of off-policy learning methods, called Experience Replay (ER) \citep{Adam:15}, \citep{Liu:16}, \citep{Zha:17}, has been shown to improve the convergence rate, computational efficiency and stability of a learning algorithm.
\item[ii)] Utilization of arbitrary pairs of states and behavior policies to be applied to the unknown system (1), enabling rich data exploration.
\item[iii)] Derivation of behavior policies which are pure PoE signals and are not superimposed on any policy, e.g. probabilistic noise or sinusoidal signals with random frequencies.
\end{enumerate}
Here, we utilize a simple yet highly effective off-policy learning scheme which exploits these important properties of off-policy learning, called Randomized Experience Replay (RER). RER proceeds with the construction of a rich offline batch of data tuples $(x_b,a_b,y_b,l_b)$, with $b=1,\ldots,N,$ and $y_b=f(x_b,a_b)$, called Experience Replay Buffer. This buffer remains fixed and is repeatedly used during every iteration $i$ for the target control policy evaluation step of Q-PI-LP and Q-VI-LP.

For each tuple in the buffer, arbitrary states $x_b$ and behavior policies $a_b$ are sampled from appropriate probability distributions. Related work involving a randomized selection of states and policies was proposed in \citep{Falsone:35}, \citep{Petretti:36} and \citep{Esfahani:37}, but only for model-based ADP. To replicate a realistic learning scenario, we assume that the functional form of the dynamics, $f$, and stage cost, $l$, are unknown but can be sampled for particular states and inputs (using, for example, a simulator or experiment). We then set $y_b=f(x_b,a_b)$ and $l_b=l(x_b,a_b)$ for $b=1,\ldots,N$. We note that the algorithms do have access to the data in the buffer, including $y_b$ and $l_b$, but not to the functions $f$ and $l$.

\subsection{PI and VI as data-driven, finite dimensional LPs}
Based on Sections $2.2$ and $2.3$, as well as the discussion on off-policy learning in the current section, we can similarly reformulate the target policy evaluation steps of PI and VI as the data-driven LP problems (14) and (15) respectively. As shown in Section $2.3$, the target policy evaluation steps of PI (10) and VI (12) involve a set of equations and recursions respectively. Therefore, the associated finite-dimensional LPs (14), (15) involve equation inequality constraints and recursion inequality constraints respectively.
\section{SIMULATION STUDIES}
\subsection{A four-dimensional open-loop unstable LTI system}
Consider a four-dimensional discrete-time LTI system,
\begin{equation*}
x_{k+1} = Ax_k+Bu_k,
\end{equation*} 
where
\begin{equation*}
A = \begin{bmatrix} 1.8 & -0.77 & 0 & 1 \\ 1 & 0 & 0 & 1 \\ 1 & 1 & 0 & 1\\ 0 & 0 & 1 &0\end{bmatrix}, B= \begin{bmatrix} 1\\ 0 \\0 \\0 \end{bmatrix}, \mathcal{X}=\mathbb{R}^{4}, \mathcal{U}=\mathbb{R}.
\end{equation*}
The open-loop eigenvalues of the system are
\begin{equation*}
[-0.4236\pm 0.6048i,0.7902,1.8569],
\end{equation*}
and therefore the system is unstable.  We use a discount factor $\gamma =0.9$ and a quadratic stage cost function defined as
$l(x_k,u_k) = x_k^{T}Ex_k+u_k^{T}Fu_k$, where $E=I_{4\times 4}$ and $F=1$. In this case, the optimal Q-function is quadratic in the states and inputs,
\begin{equation*}
Q^{\star}(x_k,u_k) = \begin{bmatrix} x_k \\ u_k\end{bmatrix}^{T}P^{\star}\begin{bmatrix} x_k \\ u_k \end{bmatrix},
\end{equation*}  
where $P^{\star}\in \mathbb{S}_{++}^{5}$. Moreover, $P^{\star}$ can be partitioned into submatrices
\begin{equation*}
P^{\star} \equiv \begin{bmatrix}P_{xx}^{\star}&P_{xu}^{\star}\\ P_{ux}^{\star} & P_{uu}^{\star}\end{bmatrix},
\end{equation*}
in the obvious way \citep{Bradtke:2}. Therefore, the optimal target policy is given by (5),
\begin{equation*}
\mu^{*}(x) = \underset{v}{\mathrm{argmin }}Q^{*}(x,v) = -(P_{uu}^{\star})^{-1}(P_{ux}^{\star}x).
\end{equation*}
To test the ability of the proposed algorithms to compute the optimal Q-function, we will consider the larger family of extended quadratic functions \citep{Barratt:33},
\begin{equation*}
\hat{\mathcal{F}}(\mathcal{X}\times \mathcal{U}) =\big\{Q(\cdot,\cdot)|Q(x,u)=\begin{bmatrix} x \\ u \end{bmatrix}^{T}\hat{P}\begin{bmatrix}x \\u \end{bmatrix} + \hat{p}\begin{bmatrix} x \\u \end{bmatrix} +\hat{s}\big\}
\end{equation*}
which includes $Q^{\star}$. Here, $\hat{P}\in \mathbb{R}^{5\times 5}$, $\hat{p} = \begin{bmatrix} \hat{p}_x& \hat{p}_u\end{bmatrix} \in \mathbb{R}^{5}$ with $\hat{p}_x \in \mathbb{R}^{4}, \hat{p}_u\in \mathbb{R}$ and $\hat{s}\in \mathbb{R}$.

Of course for the deterministic LQR problem at hand we know a priori that $p^{\star}=0_{1\times 5}$ and $s^{\star}=0$. The algorithms, however, do not know that the underlying system is linear and the cost quadratic, and we want to check whether they can guess so. Note that non-zero $p$ and $s$ may be needed for other classes of systems \citep{Wang:21}. In addition, the state-action relevance weight $c(\cdot,\cdot)$ is considered a probability measure with first and second moments $\mu_c =0_{5\times 1}$ and $\Sigma_c=I_{5\times 5}$ respectively. Therefore, the objective function of the LP problems for both Q-PI-LP and Q-VI-LP reduces to \citep{Beuchat:20},
\begin{equation*}
\int_{\mathcal{X}\times \mathcal{U}}Q(x,a)c\big(d(x,a)\big) = tr(\hat{P}\Sigma_c)+\hat{s}.
\end{equation*}
Randomized Experience Replay is implemented by constructing a buffer of $N=7000$ tuples of $(x_b,a_b,y_b,l_b),b=1\ldots,7000$, where $x_b\sim Uni(-5,5)$ and $a_b\sim \mathcal{N}(0,9)$. We initialize Q-PI-LP with the stabilizing target policy $\hat{\mu}^{0}(x)= \begin{bmatrix}-0.9 & -0.7 & -0.5 & -0.1\end{bmatrix}x$, which ensures that $J^{\hat{\mu}^{0}}(x) < \infty$ for all $x\in \mathbb{R}^{4}$. For Q-VI-LP, on the other hand, we consider the following subcases: A) $\hat{P}^{0}=I_{5\times 5}$, $\hat{p}^{0}=0_{1\times 5}$ and $\hat{s}^{0}=0$, for which we get the non-stabilizing target policy $\hat{\mu}^{0}(x)=0$ for all $x\in \mathbb{R}^{4}$, and B) $\hat{P}^{0}=I_{5\times 5}$, $\hat{p}^{0}=0_{1\times 5}$ and $\hat{s}^{0}=0$, although we apply the initial stabilizing target policy of Q-PI-LP.

Figure 1 shows the performance of the proposed Q-PI-LP and Q-VI-LP algorithms for this LTI example. The performance is assessed based on the error between two successive Q-function estimates, specifically by computing $||\hat{P}^{i}-\hat{P}^{i-1}||_{\infty}$, $||\hat{p}^{i}-\hat{p}^{i-1}||_{\infty}$ and $|\hat{s}^{i}-\hat{s}^{i-1}|$. The threshold parameter is set to $\epsilon=10^{-13}$. Q-PI-LP requires only 8 iterations to converge, while subcases A and B of Q-VI-LP require 71 iterations and 35 iterations to converge respectively. Therefore, the initialization of Q-VI-LP with a stabilizing target policy as in Q-PI-LP boosts convergence speed compared to the initialization with a non-stabilizing target policy. We then compute the error between the converged elements and the optimal ones which are computed by solving the related discrete-time algebraic Riccati equation, i.e. $||\hat{P}^{\star}-P^{\star}||_{\infty}$, $||\hat{p}^{\star}-p^{\star}||_{\infty}$ and $|\hat{s}^{\star}-s^{\star}|$, which in all cases are less or equal than $10^{-14}$. Hence, the proposed algorithms provide reliable convergence to $Q^{\star}$ for discrete-time LTI systems.
\begin{figure}[htp]
\begin{centering}$
\begin{array}{cc}
\hbox{\hspace{-1.3em}} \includegraphics[width=0.50\textwidth]{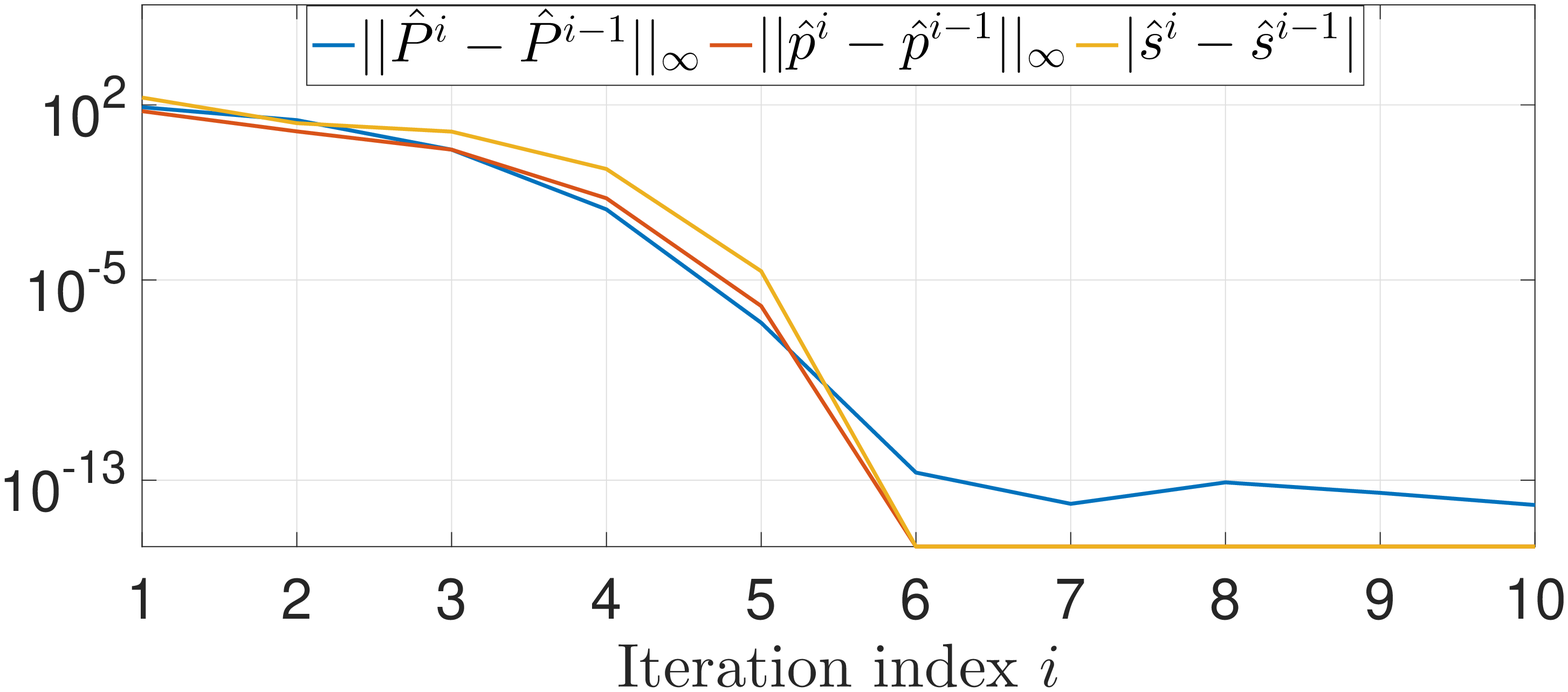} \\
\hbox{\hspace{-1.3em}} \includegraphics[width=0.50\textwidth]{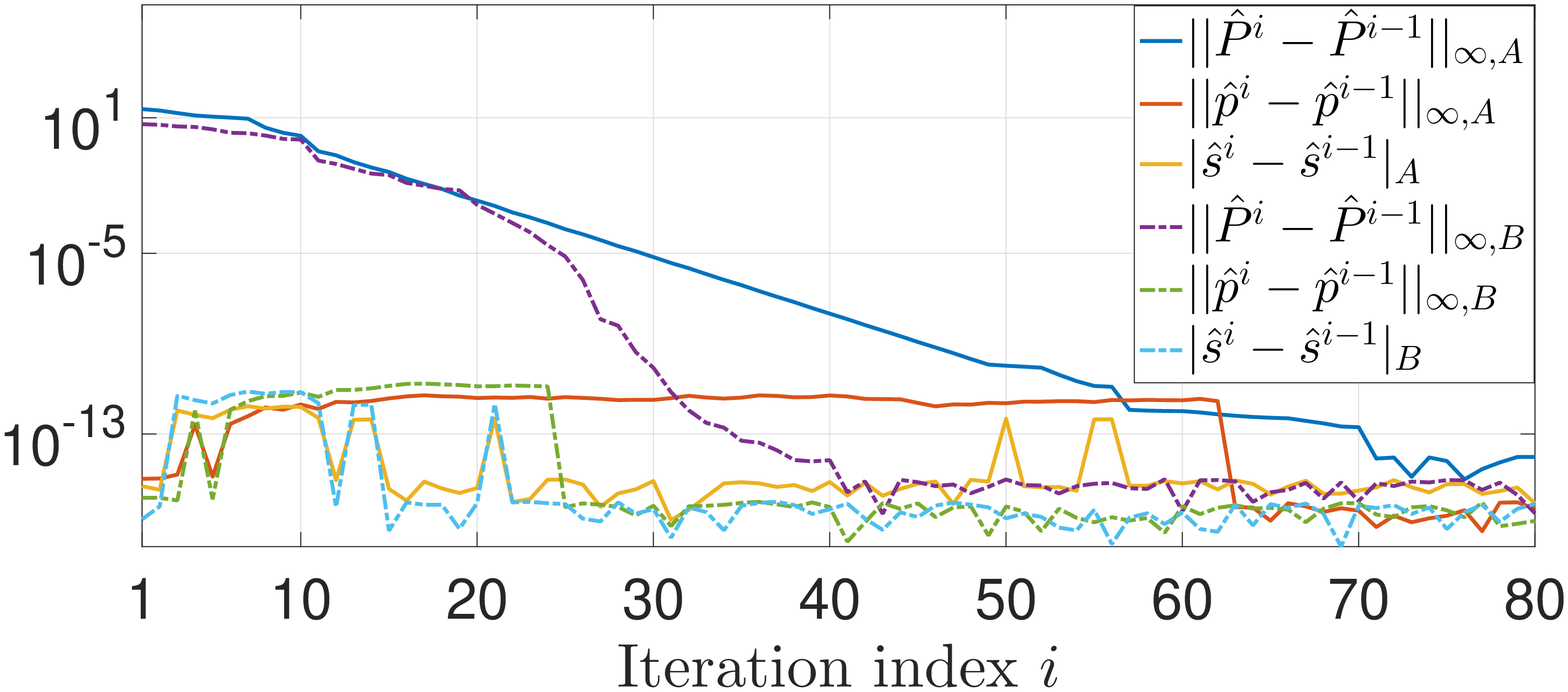}
\end{array}$
\end{centering}
\caption{Performance of Q-PI-LP (top) and Q-VI-LP (bottom) on the four-dimensional LTI example.}
\end{figure}
\subsection{A two-dimensional nonlinear system}
Consider the following two-dimensional discrete-time nonlinear system \citep{Luo:19},
\begin{equation*}
x_{k+1} = \begin{bmatrix} (x_{1,k}+x_{2,k}^{2}+u_k)\cos(x_{2,k})\\0.5(x_{1,k}^{2}+x_{2,k}+u_k)\sin(x_{2,k})\end{bmatrix},
\end{equation*}
where $\mathcal{X}=\mathbb{R}^{2}$, $\mathcal{U}=\mathbb{R}$. The following two cases are considered: i) The case of a quadratic cost function $l(x_k,u_k) = x_k^{T}Ex_k+u_k^{T}Fu_k$, where $E=I_{2\times 2}$ and $F=1$, as considered in \citep{Luo:19}, and ii) the case of a nonquadratic cost function $l(x_k,u_k) = \ln\big(x_k^{T}Ex_k+\exp(x_k^{T}Ex_k)u_k^{T}Fu_k+1\big)$, where $E=I_{2\times 2}$ and $F=1$. In both cases, the discount factor $\gamma=0.95$. Randomized Experience Replay is implemented by constructing a buffer of $N=3000$ tuples of $(x_b,a_b,y_b,l_b),b=1\ldots,3000$, where $x_b\sim Uni(-5,5)$ and $a_b\sim \mathcal{N}(0,1)$. Furthermore, the following family of quartic Q-functions is considered,
\begin{equation*}
\hat{\mathcal{F}}(\mathcal{X}\times \mathcal{U}) =\big\{Q(\cdot,\cdot)|Q(x,u)=\begin{bmatrix} x \\x^{2}\\ u \end{bmatrix}^{T}\hat{P}\begin{bmatrix}x \\x^{2}\\u \end{bmatrix}\big\},
\end{equation*} 
where
\begin{equation*}
\hat{P} \equiv \begin{bmatrix}\hat{P}_{xx}&\hat{P}_{xu}\\ \hat{P}_{ux} & \hat{P}_{uu}\end{bmatrix}\in \mathbb{R}^{5\times 5},
\end{equation*}
and all submatrices of $\hat{P}$ have identical dimensions with the ones from the simulation study in Section $4.1$.
The state-action relevance weight $c(\cdot,\cdot)$ is considered a probability measure. The additional term $x^{2}$ in the representation of the Q-function requires the first, second, third and fourth moments of $c(\cdot,\cdot)$ to appear in the objective function of the LP problems for both Q-PI-LP and Q-VI-LP. Considering the first moment as $\mu_c=0_{3\times 1}$, the objective function for both algorithms reduces to \citep{Beuchat:20},
\begin{equation*}
\int_{\mathcal{X}\times \mathcal{U}}Q(x,a)c\big(d(x,a)\big) = tr(\hat{P}_1\Sigma_c)+\hat{p}_2^{T}\phi_c +\hat{p}_3^{T}\psi_c,
\end{equation*}
where $\hat{P}_1\in \mathbb{R}^{3\times 3}$, $\hat{p}_2\in \mathbb{R}^{12}$ and $\hat{p}_3\in \mathbb{R}^{4}$ are elements of the $\hat{P}$ matrix with second, third and fourth moments given by $\Sigma_c \in \mathbb{S}^{3}$, $\phi_c \in \mathbb{R}^{12}$ and $\psi_c \in \mathbb{R}^{4}$ respectively. For this simulation study, $\Sigma_c = I_{3\times 3}$, $\phi_c = 1_{12\times 1}$ and $\psi_c=1_{4\times 1}$. Q-PI-LP is initialized with the stabilizing target policy $\hat{\mu}^{0}(x)=\begin{bmatrix}-1.5 & 0.5 & 0 & 0\end{bmatrix}\begin{bmatrix}x\\x^{2}\end{bmatrix}$ as in \citep{Luo:19}, which ensures that $J^{\hat{\mu}^{0}}(x) < \infty$ for all $x\in \mathbb{R}^{2}$.  In Q-VI-LP, on the other hand, we consider the following subcases: A) $\hat{P}^{0}=0_{5\times 5}$, where we select the initial target policy $\hat{\mu}^{0}(x)=0$ for all $x\in \mathbb{R}^{2}$, and B) $\hat{P}^{0}=0_{5\times 5}$, although we apply the above stabilizing target policy of \citep{Luo:19}. \\
Figures $2$ and $3$ show the performance of the proposed Q-PI-LP and Q-VI-LP algorithms for the cases of the quadratic and nonquadratic cost functions respectively. The performance is assessed based on the error between two successive Q-function estimates, specifically by computing the element-wise infinity norm $||\hat{P}^{i}-\hat{P}^{i-1}||_{\infty}$. The threshold parameter is set to $\epsilon=10^{-17}$. Since $Q^{\star}$ cannot be computed analytically for this simulation study, the state and control trajectories under the $\hat{Q}^{\star}$ upon convergence are also shown. 

For the case of the quadratic cost function (Figure $2$), Q-PI-LP requires 8 iterations to converge, while cases A and B of Q-VI-LP require 104 iterations and 62 iterations to converge respectively. Therefore, the initialization of Q-VI-LP with a stabilizing target policy as in Q-PI-LP boosts convergence speed compared to the simple choice of an initial zero Q-function and target policy. Both algorithms converge to the following matrix, 
\begin{equation*}
\begin{aligned}
\hat{P}^{\star} = \begin{bmatrix} 1.1154 &-0.0101 &0.0288& 0.0097 &0.6390\\ -0.0101& 1.1195& 0.0667& 0.0209& 0.0617 \\0.0288& 0.0667& 0.0023& 0.0045& -0.0305\\0.0097& 0.0209& 0.0045 &-4\cdot 10^{-4} &-0.2880\\0.6390 &0.0617 &-0.0305& -0.2880 &1.0157\end{bmatrix},
\end{aligned}
\end{equation*}
with the associated target policy,
\begin{equation*}
\hat{\mu}^{\star}(x) = \begin{bmatrix}-0.6292 & -0.0608 & 0.0301 & 0.2836\end{bmatrix}\begin{bmatrix} x\\x^{2} \end{bmatrix},
\end{equation*}
which is identical to the one computed by the state-of-art neural network-based policy gradient ADP method in \citep{Luo:19}. Furthermore, the state and control trajectories with $x_0=\begin{bmatrix}x_{1,0}\\x_{2,0}\end{bmatrix}=\begin{bmatrix}1.8\\1\end{bmatrix}$ confirm that the proposed algorithms provide reliable state-feedback regulation to the origin.
\begin{figure}[htp]
\begin{centering}$
\begin{array}{cc}
\hbox{\hspace{-1.3em}} \includegraphics[width=0.50\textwidth]{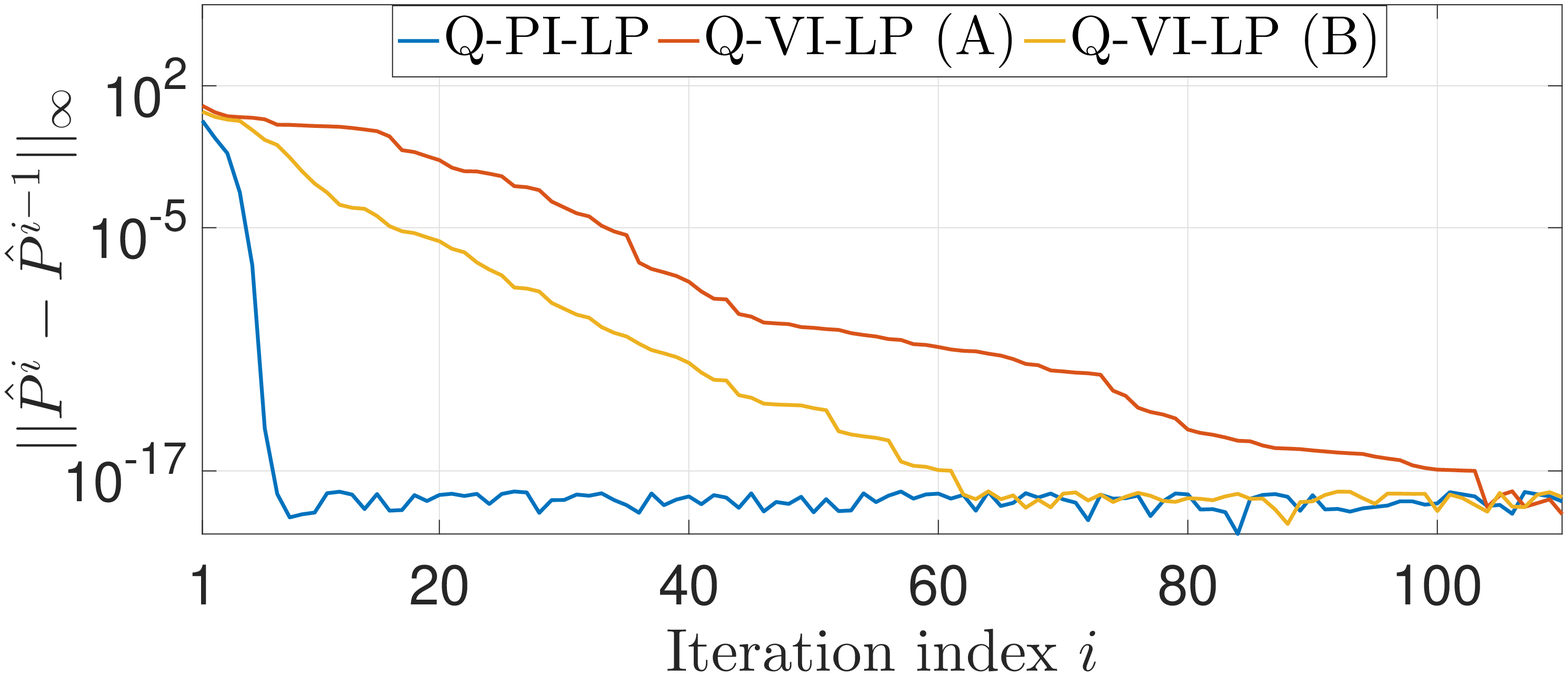} \\
\hbox{\hspace{-1.3em}} \includegraphics[width=0.50\textwidth]{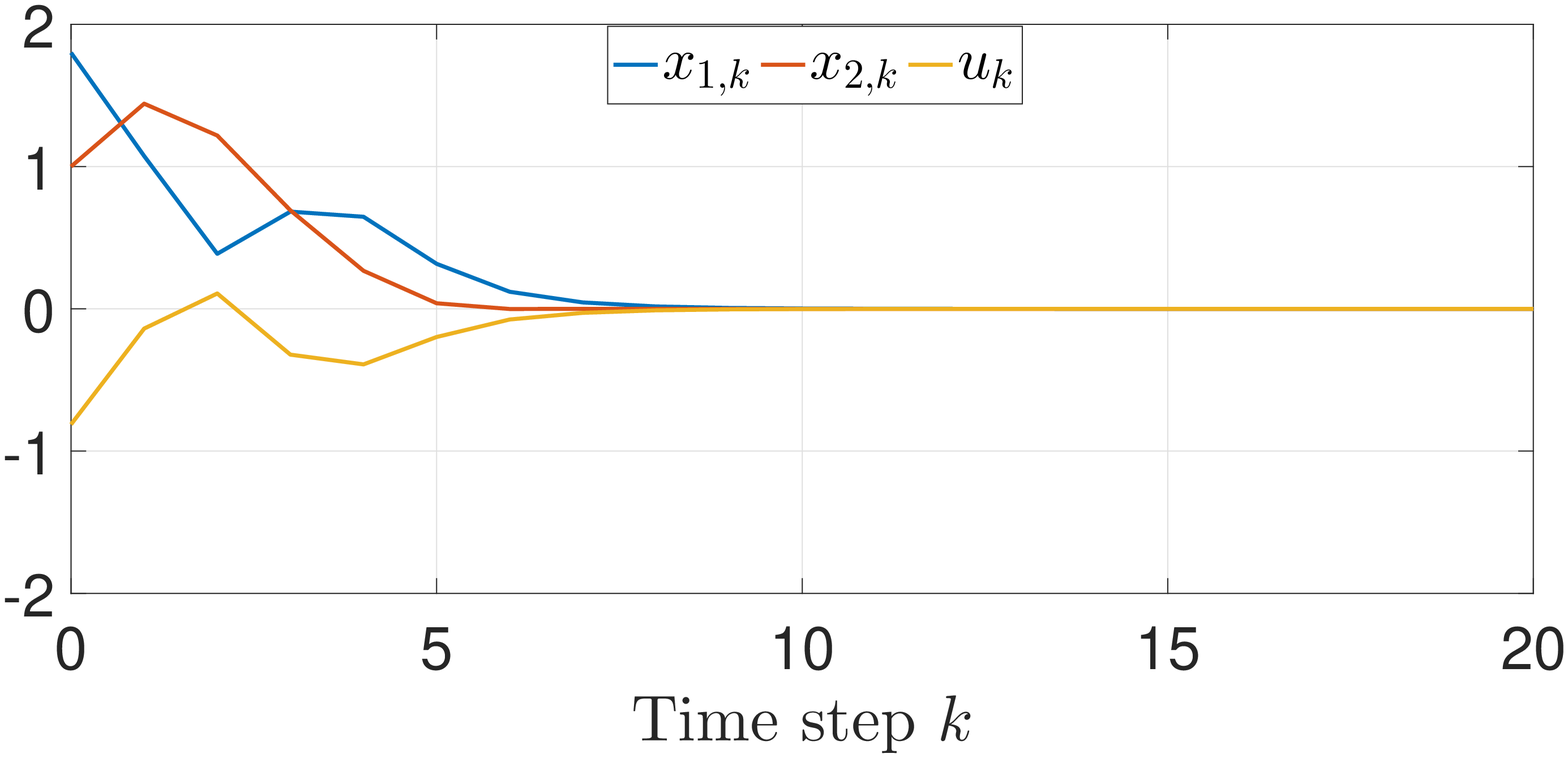}
\end{array}$
\end{centering}
\caption{Performance of Q-PI-LP and Q-VI-LP (top) and state-control trajectories under the $\hat{Q}^{\star}$ upon convergence (bottom) on the 2D nonlinear example with a quadratic cost function.}
\end{figure}

\begin{figure}[htp]                                                     
\begin{centering}$
\begin{array}{cc}
\hbox{\hspace{-1.3em}} \includegraphics[width=0.50\textwidth]{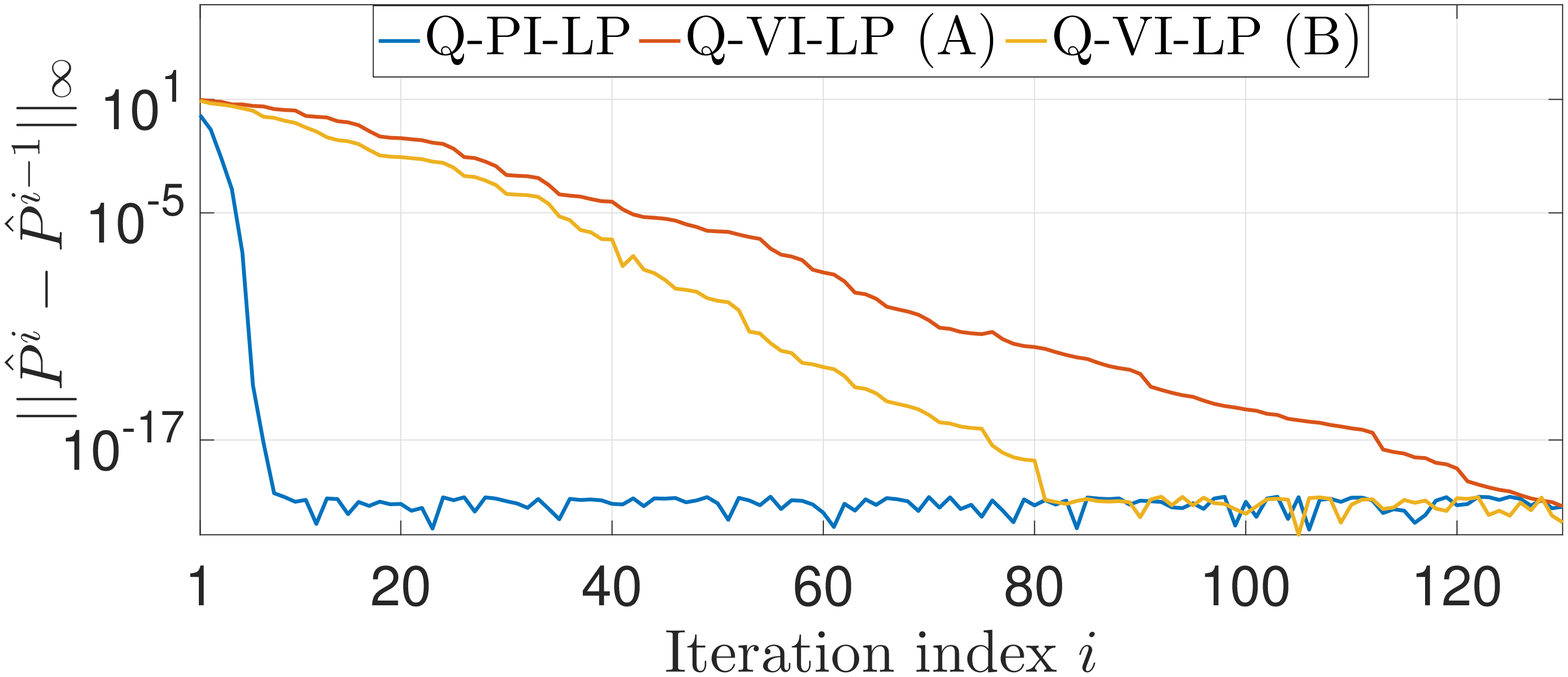} \\
\hbox{\hspace{-1.3em}} \includegraphics[width=0.50\textwidth]{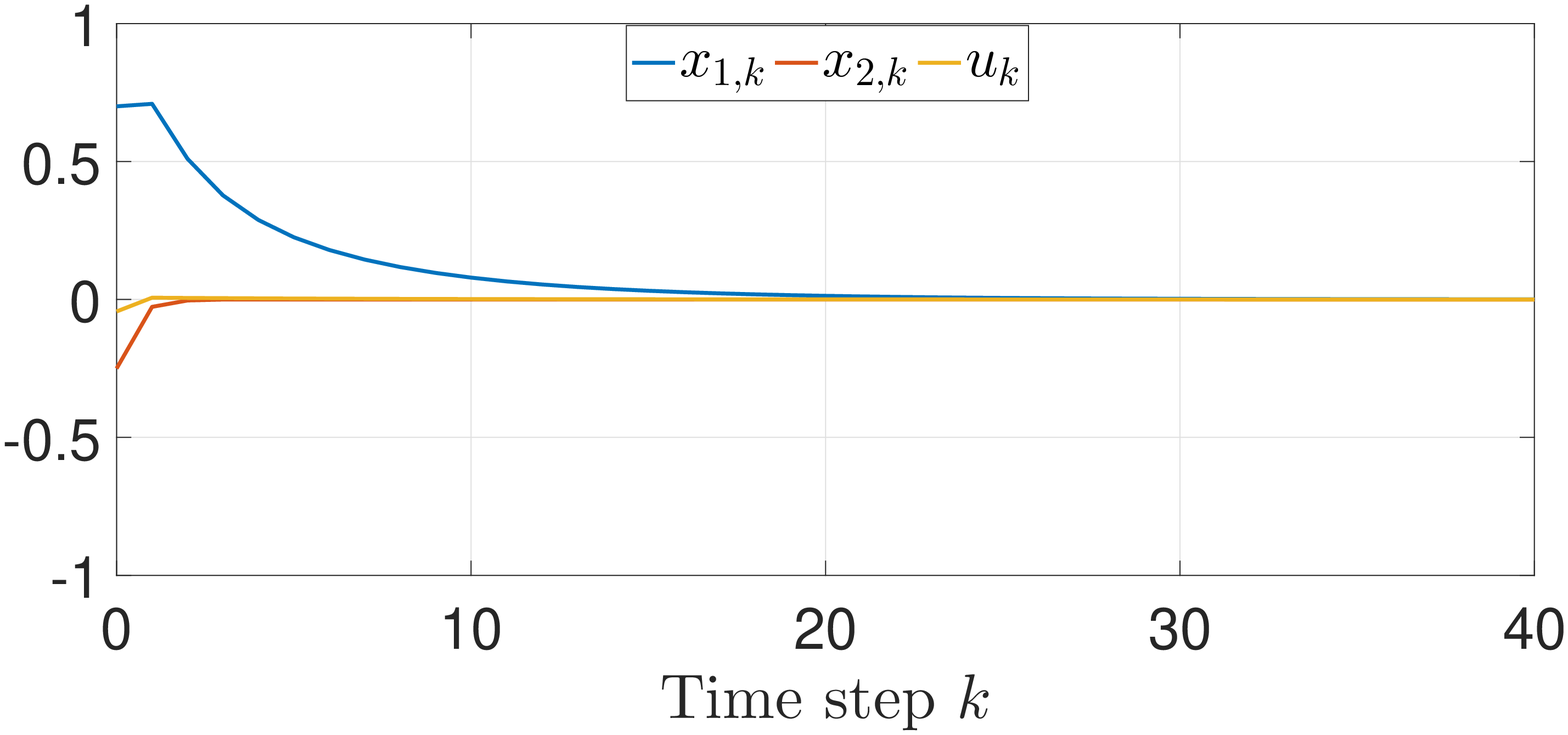}
\end{array}$
\end{centering}
\caption{Performance of Q-PI-LP and Q-VI-LP (top) and state-control trajectories under the $\hat{Q}^{\star}$ upon convergence (bottom) on the 2D nonlinear example with a nonquadratic cost function.}
\end{figure}
Finally, for the case of the nonquadratic cost function (Figure $3$), Q-PI-LP requires 9 iterations to converge, while cases A and B of Q-VI-LP require 113 iterations and 76 iterations to converge respectively. Similarly, we observe that the initialization of Q-VI-LP with a stabilizing target policy as in Q-PI-LP boosts convergence speed compared to the choice of an initial zero Q-function and target policy. Both algorithms converge to the following matrix, 
\begin{equation*}
\begin{aligned}
\hat{P}^{\star} = \begin{bmatrix} 0.6435 &0.0682 &0.0259& -0.0131 &0.0329\\ 0.0682& 0.6310& 0.1173& 0.0190& 0.1450 \\0.0259& 0.1173& 0.0146& 0.0044& 0.0451\\-0.0131& 0.0190& 0.0044 &0.0034 &0.0051\\0.0329 &0.1450 &0.0451& 0.0051 &0.2107\end{bmatrix},
\end{aligned}
\end{equation*}
with the associated target policy,
\begin{equation*}
\hat{\mu}^{\star}(x) = \begin{bmatrix}-0.1561 & -0.6881 & -0.2140 & -0.0242\end{bmatrix}\begin{bmatrix} x\\x^{2} \end{bmatrix}.
\end{equation*}
The state and control trajectories with $x_0=\begin{bmatrix}x_{1,0}\\x_{2,0}\end{bmatrix}=\begin{bmatrix}0.7\\-0.25\end{bmatrix}$ similarly show the superior capabilities of Q-PI-LP and Q-VI-LP in providing effective state-feedback regulation for nonlinear systems. Based on the theroretical discussions in Sections $2$ and $3$ and the conducted simulation studies in the current section, we observe that Q-PI-LP and Q-VI-LP inherit the monotonicity and convergence guarantees of the conventional PI and VI algorithms respectively.

We finally note that, in the data-driven on-policy PI method of \citep{Banjac:23}, an additional constraint, $\hat{P}_{uu}\succ \tau I$, with $\tau \in \mathbb{R}_{+}$ a sufficiently small constant and $I$ an identity matrix of appropriate dimensions, is added to the related LP problem, to ensure invertability of $\hat{P}_{uu}$ required for the computation of the target policy $\hat{\mu}(x)$. In our family of algorithms, this additional constraint is not necessary, since the objective function of the LP problems on Q-PI-LP and Q-VI-LP is observed to enforce the derived $\hat{P}$ to be positive definite when required, at every iteration $i$, as long as $\Sigma_c \succ 0$. 

\section{Conclusions}
In this paper, we have successfully extended the well-established model-based LP approach to ADP to the critical model-free setting. By utilizing off-policy Q-learning, RER and data-driven LP variants of PI and VI methods, we have derived novel, high-performance optimization algorithms which provide effective discounted state-feedback regulation of general unknown deterministic discrete-time systems. These successful results lead us to explore extensions of the proposed algorithms to other challenging domains, e.g. the problem of model-free optimal control with state and control constraints, robust control of unknown systems using novel data-driven $H_{\infty}$ control methods etc. An important open problem of interest is how to combine non-parametric function models (e.g. Gaussian Processes) with the proposed methods for reliable and efficient model-free optimal control of general discrete-time systems.

\bibliography{ifacconf}    
                                                   







\end{document}